\documentclass[12pt]{article}
\usepackage{amsmath}
\usepackage{amssymb}
\usepackage{epsfig}
\usepackage{euscript}
\usepackage{fancybox}
\usepackage{color}

\def\mathswitchr#1{\relax\ifmmode{\mathrm{#1}}\else$\mathrm{#1}$\fi}

\newcommand {\pslash}{\hbox{$\not\hbox{\kern-2.3pt $p$}$}}

\usepackage{cite}

\usepackage{epic}

\def\alf1{ {\alpha\over\pi} }

\begin{document}
\begin{titlepage}
\begin{flushright}
{\bf BU-HEPP-05-01 }\\
{\bf Mar., 2005}\\
\end{flushright}
 
\begin{center}
{\Large Final State of Hawking Radiation in Quantum General Relativity$^{\dagger}$
}
\end{center}

\vspace{2mm}
\begin{center}
{\bf   B.F.L. Ward}\\
\vspace{2mm}
{\em Department of Physics,\\
 Baylor University, Waco, Texas, 76798-7316, USA}\\
\end{center}

\vspace{5mm}
\begin{center}
{\bf   Abstract}
\end{center}
We use a new approach to the UV behavior of quantum general relativity,
together with some recent results from the phenomenological
asymptotic safety analysis of the theory,
to discuss the final state of the Hawking radiation for an originally
very massive black hole solution of Einstein's theory. We find that,
after the black hole evaporates to the Planck mass size, its horizon
is obviated by quantum loop effects, rendering the entire mass of the
originally massive black hole accessible to our universe.
\vspace{10mm}
\vspace{10mm}
\renewcommand{\baselinestretch}{0.1}
\footnoterule
\noindent
{\footnotesize
\begin{itemize}
\item[${\dagger}$]
Work partly supported 
by NATO Grant PST.CLG.980342.
\end{itemize}
}

\end{titlepage}

\def\Kmax{K_{\rm max}}\def\ieps{{i\epsilon}}\def\rQCD{{\rm QCD}}
\renewcommand{\theequation}{\arabic{equation}}
\font\fortssbx=cmssbx10 scaled \magstep2
\renewcommand\thepage{}
\parskip.1truein\parindent=20pt\pagenumbering{arabic}\par
Given the many successes of Einstein's classical theory of
general relativity~\cite{abs,mtw,sw1}, 
the fact that the only accepted complete treatment of quantum general
relativity, superstring theory~\cite{gsw,jp}, involves 
\footnote{ Recently, the loop quantum gravity approach~\cite{lpqg1}
has been advocated by several authors, but it has still unresolved
theoretical issues of principle, unlike the superstring theory.
Like the superstring theory, loop quantum gravity 
introduces a fundamental length
, the Planck length, as the smallest distance in the theory. This
is a basic modification of Einstein's theory.} 
many hitherto unseen degrees of freedom, some at
masses well-beyond the Planck mass, is
even more of an acute issue, as we have to wonder if such degrees of 
freedom are anything more than a mathematical artifact?
The situation is reminiscent of the old string theory~\cite{schwarz} 
of hadrons, which was ultimately superseded by the fundamental
point particle field theory of QCD~\cite{qcd1}.\par

Accordingly, in the recent literature, several authors have attempted
to apply well-tested methods from the Standard Model~\cite{qcd1,sm1} (SM) 
physics arena to quantum 
gravitational physics: in Refs.~\cite{dono1}, the famous low energy expansion technique from chiral perturbation theory for QCD has been used to address
quantum gravitational effects in the large distance regime,
in Refs.~\cite{sola} renormalization group methods in curved space-time
have been used to address astrophysical and cosmological
(low energy) effects and in Refs.
~\cite{laut,reuter2,litim} the asymptotic safety fixed-point
approach of Weinberg~\cite{wein1} has been used
to address the bad UV behavior of quantum general relativity
whereas in Refs.~\cite{bw1,bw2,bw3,bw4} the 
new resummed quantum gravity approach
(RQG) has also been used to address the bad UV behavior
of quantum general relativity (QGR). The ultimate check on these developments, which are not mutually
exclusive, will be the confrontation with experimental data. In this vein,
we focus in the following on an important issue that arises when semi-classical
arguments are applied to massive black hole solutions of Einstein's theory.\par

More precisely, Hawking~\cite{hawk1} has pointed-out that a massive black hole
emits thermal radiation with a temperature known as the Bekenstein-Hawking
temperature~\cite{hawk1,bek-hawk}. This result is well accepted by now.
This raises the question as to what is the final state of the Hawking
evaporation process? In Ref.~\cite{reuter2},  
it was shown that an originally
massive black hole emits Hawking radiation until its mass reaches
a critical mass $M_{cr}\sim M_{Pl}$, at which the 
Bekenstein-Hawking temperature vanishes and the evaporation process stops. 
Here, $M_{Pl}$ is the Planck mass, $1.22\times 10^{19}GeV$.
This would in principle
leave a Planck scale remnant as the final state of the Hawking process.\par

Specifically, in Ref.~\cite{reuter2}, the running Newton constant
was found to be
\begin{equation}
G(r)=\frac{G_Nr^3}{r^3+\tilde{\omega}G_N\left[r+\gamma G_N M\right]}
\label{rnG}
\end{equation}
for a central body of mass $M$ where $\gamma$ is a phenomenological
parameter~\cite{reuter2} satisfying $0\le\gamma\le\frac{9}{2}$,
$\tilde{\omega}=\frac{118}{15\pi}$ and $G_N$ is the Newton constant
at zero momentum transfer.
The respective lapse function in the metric class
\begin{equation}
ds^2 = f(r)dt^2-f(r)^{-1}dr^2 - r^2d\Omega^2
\end{equation}
is then taken to be
\begin{equation}
\begin{split}
f(r)&=1-\frac{2G(r)M}{r}\cr
    &= \frac{B(x)}{B(x)+2x^2}|_{x=\frac{r}{G_NM}},
\end{split}
\label{reuter1}
\end{equation}
where 
\begin{equation}
B(x)=x^3-2x^2+\Omega x+\gamma \Omega
\end{equation}
for 
\begin{equation}
\Omega=\frac{\tilde\omega}{G_NM^2}=\frac{\tilde\omega M_{Pl}^2}{M^2}.
\end{equation}
This leads to the conclusions that~\cite{reuter2} 
for $M<M_{cr}$ there is no horizon in the metric in the system and  
that for $M\downarrow M_{cr}$ the Bekenstein-Hawking temperature
vanishes, leaving a Planck scale remnant, where
\begin{equation}
M_{cr}=\left[\frac{\tilde{\omega}}{\Omega_{cr}G_N}\right]^{\frac{1}{2}}
\label{mcrit}
\end{equation}
for
\begin{equation}
\Omega_{cr} = \frac{1}{8}(9\gamma+2)\sqrt{\gamma+2}\sqrt{9\gamma+2}-\frac{27}{8}\gamma^2-\frac{9}{2}\gamma+\frac{1}{2}.
\end{equation}
For reference, we see that for the range $0<\gamma<\frac{9}{2}$ from 
Ref.~\cite{reuter2}
we have  $1>\Omega_{cr}\gtrsim .2$.

The source of these results can be seen 
to be the fixed-point behavior for the running Newton constant 
in momentum space found in
Ref.~\cite{reuter2},
\begin{equation}
G(k) =  \frac{G_N}{1+\omega G_Nk^2}
\end{equation}
where $\omega\sim 1$ depends on the the precise details of the 
IR momentum cut-off in the blocking procedure used in Ref.~\cite{reuter2}.
As we have shown in Refs.~\cite{bw1,bw2,bw3}, our RQG theory gives the same
fixed-point behavior for $G(k)$ so that we would naively conclude that
we should have the same black hole physics phenomenology as that
described above for Ref.~\cite{reuter2}. Indeed, we have shown~\cite{bw2,bw3} 
that for massive elementary particles, the classical conclusion that
they should be black holes is obviated by our rigorous quantum
loop effects, which do not contain any unknown phenomenological parameters.
We note as well that the results in Ref.~\cite{maart}, obtained
in a simple toy model using loop quantum gravity methods~\cite{lpqg1}, 
also support the
conclusion that, for masses below a critical value, black holes do not form;
the authors in Ref.~\cite{maart} are unable to specify the precise value
of this critical mass.\par

However, as we have shown in Ref.~\cite{bw1,bw2,bw3}, for elementary massive
particles, and this Bonnano-Reuter Planck scale remnant would indeed be such a
massive object with a mass smaller than many of the fundamental
excitations in the superstring theory for example, quantum loop
effects, resummed to all orders in $\kappa=\sqrt{8\pi G_N}$, 
lead to the Newton potential
\begin{equation}
\Phi_{N}(r)= -\frac{G_N M_{cr}}{r}(1-e^{-ar})
\label{newtn}
\end{equation}
where the constant $a$ depends on the masses of the 
fundamental particles in the
universe. We take here the latter particles to be those
in the SM and its extension
as suggested by the theory of electroweak symmetry breaking~\cite{ewsymb}
and the theory of grand unification~\cite{guts}. For the upper
bound on $a$ we use, we will not need to speculate about what particles
may exist beyond those in the SM
; and, for the
SM particles we use
the known rest masses~\cite{pdg2002,pdg2004}\footnote{ For the neutrinos,
we use the estimate $m_\nu\sim 3eV$~\cite{neut}.} as well as the value $m_H\cong 120GeV$ 
for the mass of the physical Higgs particle -- the latter is known to be
greater than 114.4GeV with 95\% CL~\cite{lewwg}.
\par

More precisely, when the graphs Figs. 1 and 2 are computed 
in our resummed quantum gravity
\begin{figure}
\begin{center}
\epsfig{file=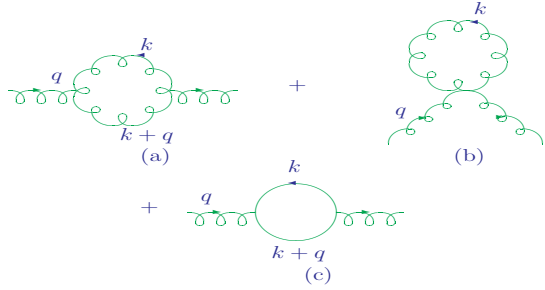,width=77mm,height=38mm}
\end{center}
\caption{\baselineskip=7mm  The graviton((a),(b)) and its ghost((c)) one-loop contributions to the graviton propagator. $q$ is the 4-momentum of the graviton.}
\label{fig1}
\end{figure}
\begin{figure}
\begin{center}
\epsfig{file=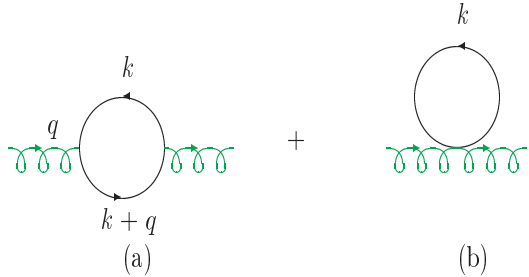,width=77mm,height=38mm}
\end{center}
\caption{\baselineskip=7mm  The scalar one-loop contribution to the
graviton propagator. $q$ is the 4-momentum of the graviton.}
\label{fig2}
\end{figure}
theory as presented in Refs.~\cite{bw1,bw2,bw3}, the coefficient $c_{2,eff}$
in eq.(12) of Ref.~\cite{bw3} becomes here, summing over the SM particles
in the presence of the
recently measured small 
cosmological constant~\cite{cosm1}, which implies the
gravitational infrared cut-off of $m_g\cong 3.1\times 10^{-33}$eV,   
\begin{equation}
c_{2,eff}=\sum_j n_j I(\lambda_c(j))
\label{ceff}
\end{equation} 
where we define~\cite{bw3} $n_j$ as the effective number of degrees of freedom
for particle $j$ and the integral $I$ is given by
\begin{equation}
I(\lambda_c)\cong \int^{\infty}_{0}dx x^3(1+x)^{-4-\lambda_c x}
\end{equation}
with the further definition $\lambda_c(j)=\frac{2m_j^2}{\pi M_{Pl}^2}$
where the value of $m_j$ is the rest mass of particle $j$
when that is nonzero. When the rest mass of particle $j$ is zero,
the value of $m_j$ turns-out to be~\cite{elswh} 
$\sqrt{2}$ times the gravitational infrared cut-off 
mass~\cite{cosm1}. We further note that, from the
exact one-loop analysis of Ref.\cite{thvelt1}, it also follows
that the value of $n_j$ for the graviton and its attendant ghost is $42$.
For $\lambda_c\rightarrow 0$, we have found the approximate representation
\begin{equation}
I(\lambda_c)\cong \ln\frac{1}{\lambda_c}-\ln\ln\frac{1}{\lambda_c}-\frac{\ln\ln\frac{1}{\lambda_c}}{\ln\frac{1}{\lambda_c}-\ln\ln\frac{1}{\lambda_c}}-\frac{11}{6}.
\end{equation} 
\par
We wish to combine our result in (\ref{newtn}) with the 
result for $G(r)$ in (\ref{rnG}) 
from Ref.~\cite{reuter2}. We do this by omitting
from the $c_{2,eff}$ the contributions from the graviton and its ghost,
as these are presumably already taken into account in $G(r)$ in (\ref{rnG}),
and by replacing $G_N$ in (\ref{newtn}) with the running result
$G(r)$ from (\ref{rnG}). 
Thus our improved Newton potential reads
\begin{equation}
\Phi_{N}(r)= -\frac{G(r) M_{cr}}{r}(1-e^{-ar}),
\label{newtnrn}
\end{equation}
where now, with 
\begin{equation}
c_{2,eff} \cong 1.41\times 10^{4}
\end{equation}
and, from eq.(8) in Ref.~\cite{bw3},
\begin{equation}
a \cong (\frac{360\pi M_{Pl}^2}{c_{2,eff}})^{\frac{1}{2}}
\end{equation}
we have that
\begin{equation}
a \cong  0.283 M_{Pl}.
\end{equation}
Since the result from Ref.~\cite{reuter2} for $G(r)$
is based on analyzing the pure Einstein theory with no matter, it only
contains the effects of pure gravity loops whereas, if we omit the
graviton and its ghost loops from our result for $c_{2,eff}$,
our result for $a$ in (\ref{newtnrn}) only contains matter loops.
Hence, there really is no double counting of effects in (\ref{newtnrn}).\par

As we have explained elsewhere~\cite{bw3}, if we use the connection
between $k$ and r that is employed in Ref.~\cite{reuter2} and restrict
our result for $c_{2,eff}$ to pure graviton and its ghost loops, we recover
the results of Ref.~\cite{reuter2} for $G(k)$ and $G(r)$ with the similar
value of the coefficient of $k^2$ in the denominator of $G(k)$, for example.
Thus, we can arrive at our result in (\ref{newtnrn}) independent
of the exact renormalization group equation (ERGE) 
arguments in Ref.~\cite{reuter2}.
A more detailed version of such an analysis will appear 
elsewhere~\cite{elswh}.\par

We also stress here that, when one uses the ERGE for a theory, one
obtains the flow of the coupling parameters in the theory. To get to 
the exact S-matrix, and derive a formula for the Newton potential
for example, one then has to employ the corresponding
improved Feynman rules for example. Thus, in the analysis in Ref.~\cite{reuter2}and in Ref.~\cite{perc}, which extends the ERGE analysis of Ref.~\cite{reuter2}
to include matter fields, one finds the results for the behavior of
the couplings at the analyzed asymptotically safe fixed point. The 
corresponding computation of the S-matrix near the fixed point
with the attendant improved running couplings is fully consistent
with our results in (\ref{newtn}),~(\ref{newtnrn})~\cite{elswh}.\par

At the critical value $M_{cr}$, the function 
$B(x)+2x^2=x^3+\Omega_{cr} x+\gamma\Omega_{cr}$ just
equals $2x^2$ at $x=x_{cr}$, 
producing there a double zero of $B(x)$ and of the lapse function
$f(r)=1+2\Phi_N$.
When we introduce our improvement into the lapse
function via $G(r)\rightarrow G(r)(1-e^{-ar})$, the effect is to reduce
the size of the coefficient of $-2x^2$ in $B(x)$ to $-2\xi x^2$
where $\xi=\xi(x)=1-e^{-aG_NM_{cr}x} < 1$ and thereby to remove 
the double zero at $x_{cr}$.
The respective monotone behaviors of the polynomials 
$x^3+\Omega_{cr} x+ \gamma\Omega_{cr}$
and  $2\xi(x_{cr})x^2$ then allow us to conclude that the lapse function
remains positive and does not vanish as $x\downarrow 0$, i.e., 
our quantum loop 
effects have obviated the horizon of the would-be Planck scale remnant
so that the entire mass of the would-be Planck scale remnant
is made accessible to our universe by our quantum loop effects.
This result holds for all choices of the parameter $\gamma$ in
the range specified by Ref.~\cite{reuter2}.

We note the nature of the way the results in Ref.~\cite{reuter2}
and our result in (\ref{newtnrn}) are to be 
combined: first one carries out the
analysis in Ref.~\cite{reuter2} and shows that the originally massive
black hole evaporates by Hawking radiation down to the
critical mass $M_{cr}$; then, in this regime of masses, the
Schwarzschild radius is in the Planck scale regime, wherein the
calculation in (\ref{newtnrn}) is applicable to show that the
horizon at $M_{cr}$ is in fact absent. One can not simply
use the result in (\ref{newtnrn}) for all values of
$M$ because it is only valid in the deep UV. Above, we have
used a step function at $x=x_{cr}$ to turn-on our
improvement for $x\le x_{cr}$.\par

This is still only a rather approximate way of combining our result in
(\ref{newtnrn}) and the result (\ref{reuter1}) of Ref.~\cite{reuter2}
and it leaves open the question as to the sensitivity of our conclusions
to the nature of the approximation.
In principle each result is a representation of the quantum loop effects
on the lapse function if we interpret these effects in terms 
their manifestations on the effective value of Newton's constant
as it has been done in Ref.~\cite{reuter2}:
\begin{equation}
f(r)=1-\frac{2G_{eff}(r)M}{r}
\end{equation} 
where we take either $G_{eff}(r)$ from (\ref{rnG}) or
we use (\ref{newtnrn}) to get $G_{eff}=G_N(1-e^{-ar})$
where now we must set $a=0.210M_{Pl}$ to reflect the effects
of pure gravity loops. The former choice is valid for
very large r, so it applies to very massive black holes
outside of their horizons whereas the latter choice should be
applicable to the deep UV at or below the Planck scale.
A better approximation is then, after the originally very massive
black hole has, via the analysis of Ref.~\cite{reuter2}, Hawking
radiated down to a size approaching the Planck size, to join the two
continuously at some intermediate value of r by determining
the outermost solution, $r_>$, of the equation
\begin{equation}
1-\frac{2G(r)M}{r}=1-\frac{2G_N(1-e^{-ar})M}{r}
\end{equation}
where $G(r)$ is given above by (\ref{rnG}),
and to use the RHS of the latter equation for $f(r)$
for $r<r_>$. For example, for $\Omega= 0.2$, we find $r_>\cong 27.1/M_{Pl}$,
so that, whereas the result (\ref{reuter1}) would give an
outer horizon at $x_+\cong 1.89$ for $\gamma=0$~\footnote{We follow Ref.~\cite{reuter2} and ask for self-consistency in the determination of
$\gamma$ and leads us to the choice $\gamma=0$ here.}, we get
$x_+\cong 1.15$ when we do this continuous combination; moreover, the
inner horizon implied by (\ref{reuter1}) at $x_-\cong 0.106$
moves to negative values of x so that it ceases to exist.
The Bekenstein-Hawking temperature
in this continuous combination
remains positive for all $x>0$ because the two equations
\begin{eqnarray}
0&=&x-2+2e^{-\frac{yx}{2}},\nonumber\\
0&=&1-ye^{-\frac{yx}{2}},
\label{sys1}
\end{eqnarray}
where $y=2\sigma/\Omega^{\frac{1}{2}}$ for 
$\sigma=(a/M_{Pl})\sqrt{\tilde\omega}$, require as well
\begin{equation}
1=ye^{-(y-1)}.
\label{sys2}
\end{equation}
The expression on the RHS of this latter equation
has a maximum for $y\ge 0$ at $y=1$ and this maximum is
just $1$, the constant on the LHS of the same equation.
The only positive 
solution to (\ref{sys2}) is then $y=1$, or $\Omega=4\sigma^2$.
This corresponds to x=0 in the (\ref{sys1}), which contradicts
the assumption therein that $x>0$. Hence, we see that
the outer horizon just approaches $x=0$ for 
$\Omega\rightarrow 4\sigma^2\equiv\Omega'_{cr}$
and that, at this point, the derivative of the lapse function
is positive. The mass $M'_{cr}$ implied by $\Omega'_{cr}$
is $2.38~M_{Pl}$. In other words, originally massive black holes
emit Hawking radiation until they reach the point $M'_{cr}\sim M_{Pl}$ 
at which their horizon vanishes, in complete agreement with our more
approximate treatment above.\par

The intriguing question is that, after reaching the final mass $M'_{cr}$,
which is now made accessible by quantum loops, how will that mass 
manifest itself? Depending on the value of its baryon number, we can expect
that there is non-zero probability for its decay into to just 
two body final states,
such as two nucleons, resulting in cosmic rays with energy $E=\frac{1}{2}M'_{cr}\cong 1.2M_{Pl}$.
More complicated decays would populate the cosmic ray spectrum with
energy $E<\frac{1}{2}M'_{cr}\cong 1.2M_{Pl}$.
Such cosmic rays
may help to explain the current data~\cite{cosmicray,westerhoff} 
on cosmic rays 
with energies exceeding $10^{19}$eV.
\par

In sum, all of the mass
of the originally very massive black hole is ultimately made accessible
to our universe by quantum loop effects. This conclusion agrees with some
recent results by Hawking~\cite{hawk2}.\par
  
\section*{Note Added}
We point out that the map given in Ref.~\cite{reuter2} for the
phenomenological distance correlation for the respective infrared cut-off
$k$ is based on standard arguments from quantum mechanics and the
parameter $\gamma$ encodes a large part of the phenomenological
aspects of that correlation. In Refs.~\cite{bw1,bw2,bw3}, the variable
$k$ is the Fourier conjugate of the position 4-vector $x$ so that
the connection from function space of $\vec k$ space to that of $\vec r$
space is given by standard Fourier transformation with no phenomenological
parameters,i.e., the result in (\ref{newtn}) does not have any sensitivity
to parameters such as $\gamma$. This underscores the correctness
of (\ref{newtnrn}) and the main conclusion we draw from it: for all choices of $\gamma$, the Planck scale remnant has its horizon
obviated by quantum loop effects.

\section*{Acknowledgments}

We thank Prof. S. Jadach for useful discussions.

\end{document}